# Review on Infrared Nanospectroscopy of Natural 2D Phyllosilicates

RAPHAELA DE OLIVEIRA,[1,2] ALISSON R. CADORE,[3] RAUL O. FREITAS,[2] INGRID D. BARCELOS[2,*]

[1]*Brazilian Synchrotron Light Laboratory (LNLS), Brazilian Center for Research in Energy and Materials (CNPEM), São Paulo, Brazil*
[2]*Physics Department, Federal University of Minas Gerais, Minas Gerais, Brazil*
[3]*Brazilian Nanotechnology National Laboratory (LNNano), Brazilian Center for Research in Energy and Materials (CNPEM), São Paulo, Brazil*
*\*Corresponding author: ingrid.barcelos@lnls.br*

**Phyllosilicates emerge as a promising class of large bandgap lamellar insulators. Their applications have been explored from fabrication of graphene-based devices to 2D heterostructures based on transition metal dicalcogenides with enhanced optical and polaritonics properties. In this review, we provide an overview on the use of IR s-SNOM for studying nano-optics and local chemistry of a variety of 2D natural phyllosilicates. Finally, we bring a brief update on applications that combine natural lamellar minerals into multifunctional nanophotonic devices driven by electrical control.**

## 1. INTRODUCTION

Natural minerals have interesting physicochemical properties and great potential for different technological applications. Phyllosilicates are a naturally abundant family of sheet silicate minerals comprising clays, micas, chlorites, serpentines, and others, which are all hydrated minerals [1,2]. They present a lamellar structure, intercalating silicon oxide tetrahedral (T) layers that form a hexagonal lattice (in ratio of 2:5 of silicon (Si) and oxygen (O) atoms) with octahedral ($O_c$) layers formed by tri and bivalent ions as central atoms, such as aluminum (Al) and magnesium (Mg), with oxygen and hydroxyl (OH) groups at the vertices (Fig. 1a). The $O_c$ layers can be of two types: dioctahedral (brucite-like), with a bivalent ion as the central atom of the octahedron, or trioctahedral (gibbsite-like), with a trivalent central atom and more compact arrangement [1,2]. The T-$O_c$-T stacking (or T-$O_c$ stacking in the case of serpentine group) of phyllosilicate minerals is weakly bound by van der Waals (vdW) force, enabling easy exfoliation down to monolayers with thicknesses ranging from 7 to 15 Å depending on the specimen [2]. Structural differences between phyllosilicate samples occur due to the different atomic substitutions in both T and $O_c$ layers. To maintain the charge balance of the structure, these atomic substitutions favor the formation of a cationic layer between the T-$O_c$-T/T-$O_c$ stacking or even the incorporation of substitutional impurities in tetrahedral and octahedral sites [1,2].

Within the large family of phyllosilicates, representative specimens such as talc, phlogopite, clinochlore, and serpentine cover the main aspects of the entire family. Talc (Fig. 1b), with chemical formula $Mg_3Si_4O_{10}(OH)_2$, is formed by the intercalation of octahedral layers of bivalent central Mg atoms with O and OH at the vertices with tetrahedral layers of silicon oxide in a T-$O_c$-T stacking. With exfoliation down to monolayer (1L) already reported in the literature [3,4], talc has a direct bandgap energy of 5.2 eV and higher dielectric constant than other insulating two-dimensional (2D) materials such as hexagonal boron nitride (hBN) [4–6]. The lamellar structure of the trioctahedral mica phlogopite (Fig. 1c) is a direct alteration of talc clay. Its chemical formula is given by $KMg_3(AlSi_3)O_{10}(OH)_2$, in which one of the four $Si^{4+}$ atoms in the tetrahedral layer is substituted by one $Al^{3+}$ atom [7,8]. This substitution results in a negatively charged structure. To maintain charge neutrality, a cationic layer of K+ atoms is formed between the T-$O_c$-T stacking. The lamellar structure of clinochlore (Fig. 1d) is formed by tetrahedral layers with one Al substitution like phlogopite. However, an $Al^{3+}$ substitution occurs in one of the three $Mg^{2+}$ sites in the $O_c$ layer of the T-$O_c$-T stacking [2,9]. Unlike talc and phlogopite, clinochlore presents an extra trioctahedral layer (brucite-like) formed purely by Mg as central atoms intercalating the T-$O_c$-T stacking, such that its chemical formula is given by $Mg_5Al(AlSi_3)O_{10}(OH)_8$. Serpentine minerals are hydrous phyllosilicates with a layered crystal structure based on overlapped T and $O_c$ layers [1,2] in a T-$O_c$ stacking. The structural form of serpentines has a tetrahedral silicate sheet with the apical oxygens shared with the corners of an Mg-rich octahedral sheet (Fig. 1e) [10,11]. Serpentine group includes antigorite and lizardite as representative specimens of high and low-temperature formations, respectively, with chemical formula $Mg_3Si_2O_5(OH)_4$ [2].

Phyllosilicates on its few-layer (FL) form have been used as substrates and encapsulating media in the manufacture of graphene-based nanodevices [5,12,13], as a capping layer protecting sensitive materials from degradation [14,15], and as substrates for transition metal dichalcogenides (TMDs) monolayers, improving the TMDs carrier mobility [16] and excitonic recombination [8,16–18]. Therefore, accessing the nanoscale properties of these materials enables the discrimination of different classes of minerals and the identification of their chemical compounds. One of the most efficient ways to access a material structure and chemistry is to probe its natural vibrational activity [10,19–22]. Thus, a systematical way to study the vibrational assignments of minerals in their FL form, which have a low response to excitation, is to use tip-enhanced techniques [23–26]. Since the small amount of material results in a low signal response in techniques with resolution at the microscale such as standard Fourier-transform infrared (FTIR) and Raman spectroscopy, scattering-type scanning near-field optical microscopy (s-SNOM) is a powerful alternative to overcome this diffraction limit [26–28] and, therefore, a tool able to characterize 2D materials composition and optical response at the nanometer scale [29]. Moreover, broadband IR imaging and spectroscopy integrated into s-SNOM setups extend the reach of the technique towards the optical characterization of 2D minerals and their vdW heterostructures to access their multiple vibrational signatures in a single experiment [24,27,29–31].

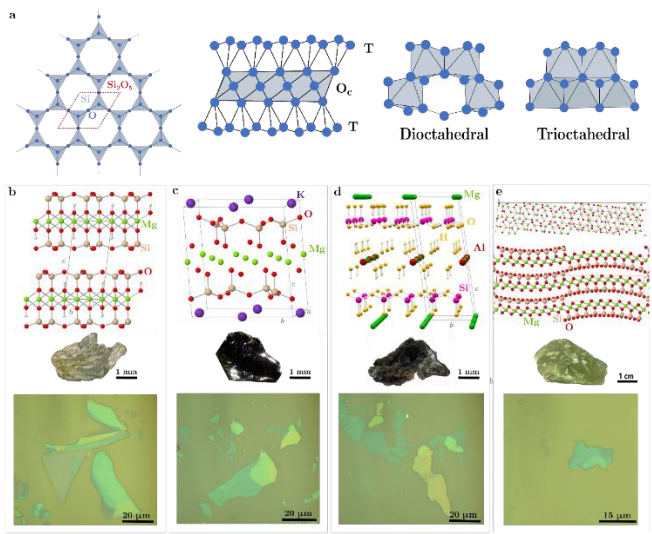

Fig. 1. The general structure of phyllosilicate minerals. (a) The panel shows a hexagonal lattice (top view) formed by Si tetrahedra where three O at the corners are shared between Si atoms and the remaining O atom points outwards. This tetrahedral (T) layer is the basis of the phyllosilicate minerals, forming an alternating stacking between T and octahedral ($O_c$) layers, which can be dioctahedral or trioctahedral (lateral view). (b) Talc, (c) phlogopite (d) clinochlore, and (e) antigorite are examples of each of the four subgroups of phyllosilicates: clays, micas, chlorites, and serpentines; respectively. The panels show the atomic structure of each mineral (top), followed by a photograph of their bulk form (middle) and an optical microscopy image (bottom) of several few-layer flakes obtained by mechanical exfoliation onto $SiO_2$/Si (300 nm thick $SiO_2$) substrates.

In the presented scope, this Review article displays how infrared (IR) nanospectroscopy is helping to unveil fundamental optical properties of the 2D natural phyllosilicates and their vdW heterostructures. Regarding investigations in the FL form of phyllosilicates, we will discuss talc, muscovite, phlogopite, and clinochlore as representative specimens of clays, dioctahedral micas, trioctahedral micas and chlorites, respectively, complemented by an antigorite-serpentine study case, covering the main aspects of the entire phyllosilicate family. Further, we briefly introduce the synchrotron s-SNOM experiment, commonly named synchrotron infrared nanospectroscopy (SINS), as an ultimate tool to study the vibrational signatures of these minerals in their FL form at the nanoscale. Hence, this Review brings several examples of the use of SINS technique for assigning the vibrational identity of FL-phyllosilicates. Finally, we revisit advances on applications comprising different natural minerals, graphene and phyllosilicate layers, for creating ultrathin vdW heterostructures. Further, we highlight investigations of graphene polaritonic activity in these natural crystals, the electrical modulation of their fundamental properties and the appearance of hybrid plasmon-phonon modes.

## 2. RAMAN SPECTROSCOPY APPLIED TO NATURALLY OCCURING PHYLLOSILICATES

All geological specimens of phyllosilicates shown in Fig. 1 were obtained from Minas Gerais (Brazil). From bulk to ultrathin flakes, the standard scotch tape exfoliation method was carried out to release and transfer 1L and FL-phyllosilicates atop two different substrates. The produced flakes were pre-characterized by optical microscopy as shown in Fig. 1, and the different colors of the flakes correspond to different thicknesses.

To investigate the vibrational properties of phyllosilicate materials, both non-destructive spectroscopy techniques can be used: Raman and Infrared [19,32–35]. Fig. 2a-d displays representative Raman spectra for the selected phyllosilicates: talc, phlogopite, clinochlore, and serpentine, respectively. Fig. 2 compares the Raman features of the bulk crystals with respect to exfoliated flakes (thicknesses larger than 200 nm) onto $SiO_2$/Si and Au substrates.

In Fig. 2a, Raman spectrum displays several modes located in the spectral window from 75 to 1350 cm$^{-1}$ and from 3550 to 3800 cm$^{-1}$, e.g., fundamental modes in silicates. The Raman peaks at about 194.5, 362.1, 676.2 and 1050.5 cm$^{-1}$ are ascribed to the fundamental vibrations of silicates in talc. The spectrum is dominated by rigid translation and rotation of the T-$O_c$-T modules (80–250 cm$^{-1}$), lattice modes involving Mg–O bonds and tetrahedral bending (250–650 cm$^{-1}$), strong bending and stretching modes of Si–O–Si bridges, around 670 cm$^{-1}$ and above 1000 cm$^{-1}$, respectively [19,36]. Moreover, Ref. [24] has used symmetry analysis to identify all Raman-active modes of talc and showed that a symmetry crossover occurs from 1L to 2L thickness, attributed to the stacking of adjacent layers, changing from $C_{2h}^3$ to $C_i^1$ space group, respectively.

In Fig. 2b, the phlogopite-Raman spectrum displays several modes. The spectrum collected is close to the Raman-active fundamental modes theoretically predicted [37]. The spectrum shows two characteristic peaks at 197 and 683 cm$^{-1}$ corresponding to the vibration of $MgO_6$ octahedra and $TO_4$ (T = Si, Al) tetrahedra, respectively, besides several other modes related to extra common modes for silicates (stretching and bending Si–O–Si vibrations, besides translational (Mg,Al,Fe)-O vibrations in the $O_c$ layers). Moreover, the position and shape of some Raman modes of phlogopite are strongly affected by changes in impurity concentration. For instance, the presence of Fe ions induces a peak splitting of the Si–O–Si (700-800 cm$^{-1}$) modes and the appearance of the strong peak in the 500-550 cm$^{-1}$ range [8].

The Raman spectrum of a bulk-clinochlore sample shown in Fig. 2c is in close agreement with the spectrum theoretically calculated [38]. The spectrum features 10 strong and well-defined modes. The peaks around 1089 and 1059 cm$^{-1}$ are attributed to antisymmetric Si-O-Si stretching modes, while the symmetric Si-O-Si stretching appears as strong bands at 687 and 553 cm$^{-1}$. The peaks at 459 and 440 cm$^{-1}$ are probably assigned to librational OH and bending Si-O-Si modes. The strong band at 353 cm$^{-1}$ has been assigned to the bending vibration of Si-O and its intensity decreases with the Si substitution by Al. The peak at 283 cm$^{-1}$ is probably related to movements of the T sheet, whereas the sharp peak at 203 cm$^{-1}$ was ascribed to vibrations of the octahedrons. Finally, the peak at 125 cm$^{-1}$ is regarded as translational Si-O-Si modes. It is worth mentioning that the two other weaker peaks observed at around 106 and 158 cm$^{-1}$ are related to bending Si-O and to a combination of the out-of-plane vibration of the Mg ions with librations of the OH groups in the brucite-type interlayer, respectively [9,19,38].

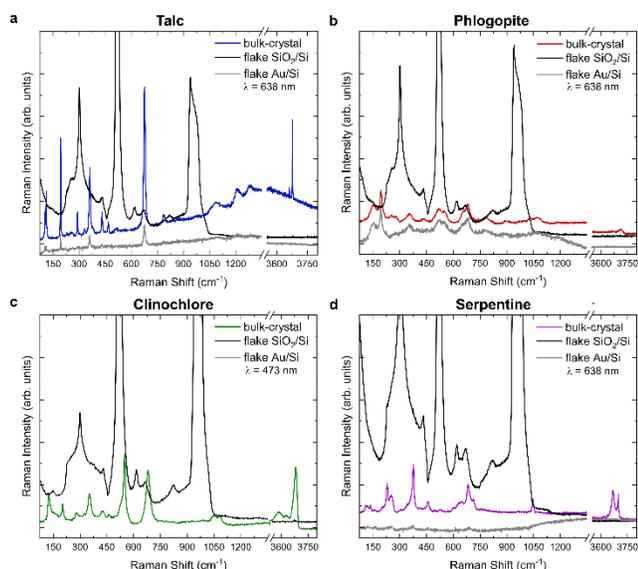

Fig. 2. Raman Spectroscopy. Raman spectrum excited at 638 nm and 473 nm of the bulk-crystal (blue, red, green, and purple curves) and exfoliated flakes onto SiO$_2$/Si (black curves) and Au (gray curves) substrates of (a) talc, (b) phlogopite, (c) clinochlore, and (d) serpentine, respectively.

Fig. 2d plots the Raman spectrum of a natural serpentine sample [11]. In this case, the Raman spectra present bands corresponding to the inner vibrational modes of the lattice and to SiO$_4$ vibrations [39]. One can observe a band occurring at 1045 cm$^{-1}$ that corresponds to the antisymmetric stretching modes of the Si-O-Si. An intense band at 680 cm$^{-1}$ due to symmetric stretching modes of the Si-O-Si linkages and a band at 630 cm$^{-1}$ ascribed to HO-Mg-OH translation modes. There is a band at 375 cm$^{-1}$ due to bending modes of the SiO$_4$ tetrahedra and a band at 226 cm$^{-1}$, where vibrations of O-HO groups are expected (O is the non-bridging oxygen atom of the tetrahedron and H is the hydrogen atom of the OH group tilted toward an octahedral cation vacancy) [39].

As presented in Fig. 2, it is extremely challenging to access the Raman signatures of isolated FL-phyllosilicate flakes via far-field analysis onto SiO$_2$/Si substrates (black curves). Due to the diffraction-limited size of the micro-Raman probe, the experiment sensitivity or signal-to-noise ratio (SNR) is mainly reduced, and the vibrational response becomes dominated by the supporting substrate. This behavior was also observed for muscovite [40] and reported elsewhere [4,41,42]. Nevertheless, for FL-talc flakes onto HOPG substrates, Refs. [4,42] demonstrated that only the three most intense modes are observed around 194.8, 676.2, and 3678.8 cm$^{-1}$. Moreover, by having the phyllosilicate flakes exfoliated onto Au substrates (gray curves in Fig. 2), we have observed very few Raman modes for each material. Thus, exemplifying the difficulty of assessing the vibrational fingerprints of FL-phyllosilicates by Raman spectroscopy. In that perspective, sub-diffractional analytical modalities such as s-SNOM and SINS are key tools. With a tip-driven lateral resolution (∼25nm), high spectral irradiance throughout mid-IR to THz ranges and penetration depth of a few tens of nanometers, SINS is now established as a decisive modality to study atomic layered materials and, therefore, a robust alternative for opto-vibrational characterization of FL-phyllosilicates.

## 3. SYNCHROTON INFRARED NANOSPECTROSCOPY AND NANOIMAGING

Standard FTIR spectroscopy is a well-established technique in the study of natural crystals [43]. This technique enables in-situ and non-destructive chemical identification with a micrometric resolution of organic matter and mineral phases in geological samples [43,44]. The micro-FTIR beam size using a conventional IR source can reach a lower limit of tens of micrometers and be further reduced to 3–5 μm using a synchrotron radiation source [43]. The use of synchrotron IR radiation not only improves the spatial resolution of the technique but also allows the investigation of very low concentrations of chemical species due to minimal signal loss [45]. As an example, high-resolution synchrotron micro-FTIR imaging of low-concentration fluid inclusions, such as water and CO$_2$ gas, were reported in Earth science studies [45]. However, the spatial resolution of FTIR spectroscopy even using synchrotron radiation is optically limited by diffraction, which intrinsically scales with the radiation wavelength.

The SINS development, which combines a broadband synchrotron IR source with s-SNOM technique, has overcome the diffraction limit of conventional optics to provide spatial resolution down to tens of nanometers [27,46]. Comparative studies using conventional- and nano-FTIR techniques allow a direct association of FTIR absorption bands with the imaginary part (phase) of SINS complex signal response [46], and offer an approach to extracting dielectric function of nanomaterials [47]. Particularly in the study of phyllosilicates, two specificities of this nanoprobe technique to the detriment of micro-FTIR were reported for the case of talc crystals [24,48], wherein the SINS technique not only provides nanometric resolution for vibrational analysis of natural crystals but also enables the study of confinement effects by varying the number of crystal layers. First, the most prominent bulk-talc mode in the SINS phase spectrum at 534.1 cm$^{–1}$ corresponds to essentially out-of-plane atomic motions, revealing a preferential sensitivity to out-of-plane vibrations probed by SINS [24]. Second, a SINS scattering intensity modulation according to the number of talc layers [24,48]. Thus, it acts as a directional probe by preferentially accessing vibrational modes oriented parallel to the electromagnetic mode of the strongly polarized near-field dipole at the SINS-tip.

s-SNOM is a well-known analytical modality that combines atomic force microscopy (AFM) optical microscopy for achieving nanoscale spatial resolution for vibrational analysis [49–51]. Among several advantages, s-SNOM allows measuring morphology and optical responses of materials surface simultaneously, enabling structural and dielectric analysis in a single experiment (Fig. 3) [29]. From the optical perspective, s-SNOM is a scattering-based modality mediated by a metallic nanoantenna (AFM tip), whose free charges are polarized by the incident light enabling strong field confinement at the tip apex. In this configuration, the confinement volume is no longer dependent on the wavelength but on the antenna shape, overcoming the diffraction limit even for extremely long wavelengths [52]. The AFM operates in semi-contact (tapping) mode, whose oscillation is set near the natural cantilever resonance frequency Ω. Far-field background suppression is achieved by applying higher harmonics (nΩ for n>2) demodulation with a lock-in amplifier [53].

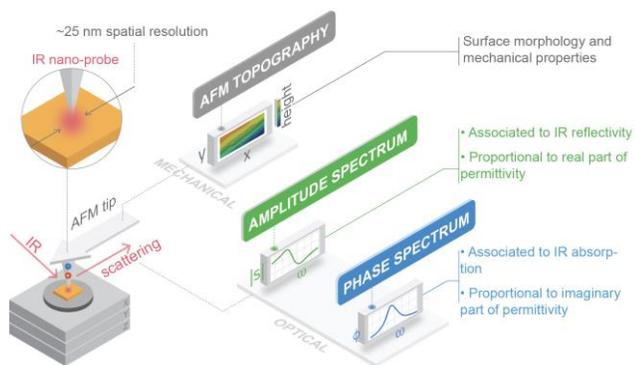

Fig. 1. s-SNOM concept scheme. As a tip-enhanced modality technique, s-SNOM overcomes the diffraction limit via field confinement at the apex of a nanoantenna (metallic AFM tip). By combining IR and AFM microscopies, the technique reaches nanometric lateral resolution in an experiment that retrieves both morphology and optical responses, simultaneously. By applying asymmetric interferometry, s-SNOM allows complex detection of the scattering radiation. Therefore, with amplitude and phase of the local scattering, s-SNOM can access IR reflectivity and absorption phenomena of any vibrational active material. Figure adapted from Ref. [54].

A unique feature of s-SNOM is its ability to measure complex scattering properties of the materials, giving the complete nano-resolved ultra-broadband near-field response, $\xi_n(\omega) = S_n(\omega)e^{i\varphi_n(\omega)}$, where $S_n(\omega)$ and $\varphi_n(\omega)$, are the s-SNOM amplitude and phase, respectively, demodulated at the $n^{th}$ tip harmonic. All SINS data here refers to n = 2. In the case of SINS measurements (i.e., broadband s-SNOM spectroscopy), we record both amplitude and phase spectra simultaneously. Note that $\varphi_n(\omega)$ is acquired owing to the lock-in based detection in an asymmetric interferometry scheme [53]. In such a configuration, phase delays driven by the sample's permittivity are separated from the interferometer modulation, allowing phase contrast imaging [55] and spectroscopy at the nanoscale [56,57]. Like classical FTIR spectroscopy, a reference spectrum needs to be collected to separate sample's response from the environmental influence, source emission coverage and the detector sensitivity. Therefore, it is common practice in SINS to measure a flat responsive material (such as Au or Si) to be used a reference complex spectrum ($\xi_{n,ref}(\omega) = S_{n,ref}(\omega)e^{i\varphi_{n,ref}(\omega)}$). Hence, a normalized spectrum is achieved from the complex ratio $\xi_n(\omega)/\xi_{n,ref}(\omega)$ that yields normalized amplitude ($S_n(\omega)/S_{n,ref}(\omega)$) and phase ($\varphi_n(\omega) - \varphi_{n,ref}(\omega)$).

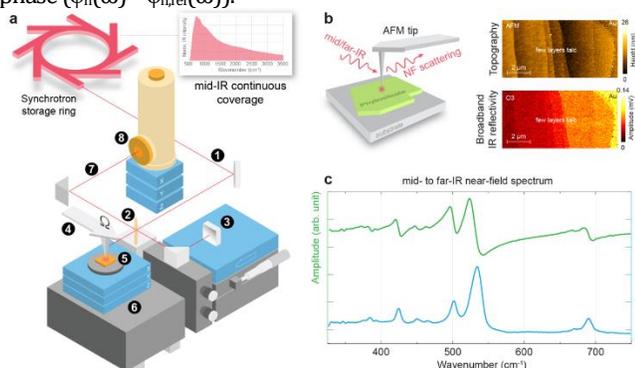

Fig. 2. Synchrotron Infrared Nanospectroscopy technique. (a) Based in the s-SNOM concept, SINS allows expanding the reach of nanoscale IR analysis powered by the ultrabroadband and continuous energy coverage of synchrotron sources. In SINS, a broadband IR beam (1) produced by an electron storage ring illuminates a metallic AFM tip (4) producing a dense field confinement at its apex. Field confinement is restricted to the tip size, therefore, no longer depends on the wavelength (sub-diffractional confinement). For demultiplexing the broadband response, the beam is divided by a beamsplitter (2) into two arms: a reference arm for phase modulation (3) and the tip-sample stage (5), that is mounted in one of the arms of an asymmetric Michelson interferometer. Backscattered, the beam is recombined by the beamsplitter and then pointed (7) towards an IR detector (8). (b) Example of SINS broadband image of a phyllosilicate crystal (left) that access both morphology (center) and IR reflectivity (right), simultaneously. (c) Typical SINS spectrum of a talc flake. The extended coverage of the synchrotron allows exploring mid- to far-IR vibrational responses in a single experiment with excellent SNR. Fig. (a) adapted from [54].

The s-SNOM data is strongly connected to the complex material's permittivity. Therefore, the amplitude SINS spectra are proportional to the real part of the permittivity, a quantity related to the reflectivity power of the material. On the other hand, the s-SNOM phase is proportional to the imaginary part of the permittivity and, hence, provides information on the absorption power of the material (Fig. 3). For instance, Fig. 4 shows SINS amplitude and phase spectra of a talc flake. The lineshape of the SINS data qualitatively resembles a driven damped harmonic oscillator resonance, a lineshape also expected for the permittivity at resonances. Moreover, scientific questions from many fields were benefited by the complex optical sensitivity of s-SNOM combined to its extreme spatial resolution including 2Ds polaritonics [29,58], 2Ds twistoptics [59], biological applications [60,61], fundamental condensed matter physics and many other fields.

In early 2010's, F. Huth et al. presented a proof-of-principle of s-SNOM operating with a broadband IR thermal source [62]. This demonstration was seminal for advancing this ultramicroscopy modality towards chemical analysis as it could operate as classical IR spectrometer, however, with sub-diffraction resolution (nano-FTIR). Despite the successful demonstration, black body sources are far from ideal for s-SNOM as they are not polarized and usually offer modest spectral irradiance. Hence, alternative broadband sources were adapted to s-SNOM, such as plasma sources [26,63,64], lasers based on difference frequency generation (DFG) [56,57] and synchrotron IR radiation [65]. In case of synchrotron radiation (SR) sources, these are highly brilliant compared to thermal sources (500 to 1000 times more brilliant) and offer extreme IR coverage, spanning from THz to near-IR [66]. Therefore, the enhanced spectral irradiance throughout the entire IR range of SR sources enabled ultrabroadband IR nanoscale spectroscopy, commonly named synchrotron infrared nanospectroscopy [27,67,68]. In this modality, a SR IR beam serves an asymmetric Michelson interferometer where the tip-sample stage is mounted inside one of the interferometric arms, Fig. 4a. The IR beam is optically processed by the beamline primary optics to enter the interferometer as a collimated or parallel beam (Fig. 4a,1). The beam is then divided by a beamsplitter (Fig. 4a,2), propagating to a reference arm (Fig. 4a,3), whose travel produces the phase modulation in the interferogram. The other beam portion is propagated to the tip-sample stage (Fig. 4a,4), where the beam is focused onto the AFM tip (Fig. 4a,5). The piezoelectric sample scanner (Fig. 4a,6) enables localized point spectroscopy or near-field imaging. Back-reflected from reference and tip-sample arms, the beam is recombined (Fig. 4a,7) after the beamsplitter and then propagated to a single-point IR detector (Fig. 4a,8). This configuration produces broadband IR interferograms that are demultiplexed by Fourier post-processing. The result is a broadband spectrum (e.g., top-right inset panel in Fig. 4a) collected from a sample area comparable to the tip radius (~25x25 nm$^2$).

In the perspective of vibrational identity of FL-phyllosilicates, SINS was a decisive tool for a first approach on the IR signatures of naturally occurring atomic-thin talc flakes. The wide coverage of the synchrotron allowed simultaneous monitoring of in-plane and out-of-plane stretching and bending modes in a single SINS experiment [24,48]. As all these bands were measured in a single spectrum and the study unveiled a direct connection between the number of talc layers and the Si-O stretching modes. Typically, the SINS experiment starts with the imaging of the phyllosilicate flake that is mounted on a flat substrate (Fig. 4b, left panel). The AFM scan reveals the surface morphology of the flake, delimiting the different atomic terraces in a range of few microns area (Fig. 4b, right panel). As the tip is Illuminated by the broadband IR from the synchrotron, the confined fields at the apex interact with the sample surface during the AFM mapping, therefore, the IR scattering is collected simultaneously. Fig. 4b right panel also shows a common broadband IR amplitude map that highlights the IR reflecting power of the flake in the whole IR range covered by the detector. The Au substrate has a superior reflectance compared to the talc flake and, consequently, the thinner plateaus of the flake have higher reflectance than the thicker ones. Once spatially mapped, a point spectrum is taken at specific locations of the flake to retrieve the spectral response of a certain number of talc layers. For instance, Fig. 4c shows a representative SINS spectrum of a talc flake acquired at the Advanced Light Source that offers a SINS setup operating from mid- to far-IR [24,48]. So far, SINS has demonstrated a great potential to study phyllosilicates as it has enhanced sensitivity to Si-O bands a simultaneous coverage of a large portion of the IR spectrum.

## 4. NANO-FTIR AND s-SNOM APPLIED TO NATURALLY OCCURING PHYLLOSILICATES

### A. Muscovite – Dioctahedral Mica

Muscovite is a dioctahedral mica, with chemical formula $KAl_2(Al,Si_3)O_{10}(OH)_2$, and it is among the earliest known vdWs materials that can be exfoliated in high quality to large sizes with atomic flatness, as well as down to 1L [40,69]. Muscovite mica is known to be a large bandgap material (5.1 eV) [1,69] with several industrial applications. Since its surface is atomically flat, muscovite mica was considered an excellent substrate for materials used in flexible devices and optoelectronics applications with other 2D materials in vdW heterostructures [13,40,70–73].

Ref. [40] has pointed out the difficulty to probe the vibrational modes of ultrathin muscovite flakes. So, Fali and colleagues have used a combination of mid-IR s-SNOM and nano-FTIR techniques to perform near-field spectroscopy and imaging of 2D crystal muscovite mica down to the 1L limit on different substrates [74]. Fig. 5a brings a representative AFM topography image, followed by narrow-band s-SNOM amplitude and corresponding phase images of muscovite mica exfoliated on a Si substrate taken at selected excitation laser frequencies (1030, 1075, and 1175 cm$^{-1}$). The authors claim that although the excitation frequency of 1075 cm$^{-1}$ is of particular interest since it is in the hyperbolic polaritonic range (1050–1130 cm$^{-1}$) of mica, the amplitude/phase images did not display the characteristic interference fringes (due to large losses) that are the hallmark of polaritonic propagating modes. Nevertheless, they have also performed nano-FTIR measurements to understand the full wavelength and thickness-dependent IR optical features.

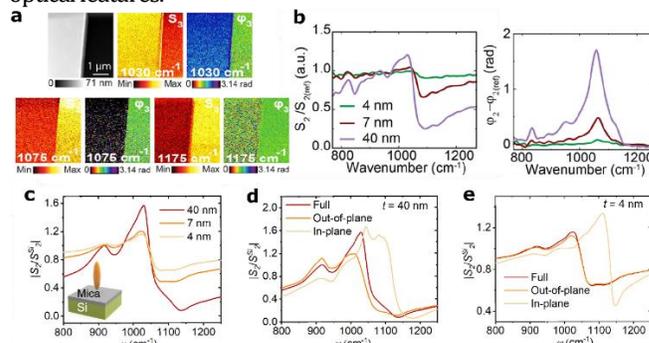

Fig. 3. Infrared dielectric properties of muscovite mica. (a) AFM topography image, third harmonic near-field amplitude, and phase images at three different frequencies of 55 nm mica flake on Si substrate. (b) Near-field experimental nano-FTIR amplitude S$_2$ and phase φ$_2$ spectra at different mica thicknesses. (c) Simulated S$_2$ amplitude spectra of 40, 7, and 4 nm mica slab on Si substrate. (d) Simulated S$_2$ amplitude spectra of 40 nm mica on Si substrate considering the full dielectric tensor, only the out-of-plane dielectric function, and only the in-plane dielectric function. (e) Similar to (b) but for 4 nm mica. Fig. (a-b) adapted from Ref. [74] Copyright 2019: The American Chemical Society, whereas Fig. (c-e) adapted from Ref. [75].

Fig. 5b then brings the experimental normalized amplitude and phase spectra, respectively. Both amplitude and phase spectra show similar appearances for different layer thicknesses; however, the signal level becomes progressively smaller with decreasing thickness due to a smaller probed volume. The broad and strong band between 980 and 1200 cm$^{-1}$ centered around 1080 cm$^{-1}$ was assigned to stretching vibrations of Si–O; the shoulder at 831 cm$^{-1}$ to the stretching Al–O mode, whereas the band near 920 cm$^{-1}$ was assigned as a mixed modes arising from Al–OH stretch, Al–O–Al vibration, and Si–O–Si and Si–O–Al stretching vibrational modes. Finally, the authors claimed that their measurements enabled the extraction of the dielectric function in this spectral range, highlighting that the IR dielectric permittivity of mica along the ordinary and extraordinary directions have opposite signs in the 920–1130 cm$^{-1}$ range, thus, implying regions of both Type I and Type II hyperbolic behavior.

After this experimental study, Chen and coworkers attempted to obtain the in-plane and out-of-plane dielectric functions of mica by COMSOL simulations [75]. Fig. 5c plots the simulated S$_2$ amplitude spectra for 40, 7, and 4 nm mica thick on Si substrate. Next, the authors examined the in-plane and out-of-plane responses individually. Fig. 5d-e shows the calculated spectra of 40 and 4 nm mica thick on Si substrate, respectively. The spectra were obtained using the full dielectric tensor, only the in-plane dielectric function, and only the out-of-plane dielectric function. The authors then observed that for thicker samples although only considering the out-of-plane component of the dielectric tensor yields a similar result compared to considering the full tensor, the effect of the in-plane response was non-negligible. Whereas, for thinner films, they claim that the out-of-plane response is indeed dominant. Finally, we must comment that such point-spectra analysis as a function of mica thickness could be done by SINS experiments, thus corroborating or not the findings claimed in such theoretical work.

### B. Phlogopite – Trioctahedral Micas

Phlogopite is a common phyllosilicate within the trioctahedral mica group with chemical formula $KMg_3(AlSi_3)O_{10}(OH)_2$ from which a solid-solution series can be formed in relation to Fe concentration arising as main impurity [1,2,8]. Exfoliated phlogopite-series flakes have been shown and investigated by scanning microscopy techniques techniques [76–78], as well as combined with other 2D materials [8]. The crystals show a monoclinic layered structure (with $C_{2/m}$ symmetry) with a T-$O_c$-T stacking. Cadore and co-workers have investigated the influence of Fe impurities on the optoelectronic properties of phlogopite and revealed that the incorporation of Fe ions in the phlogopite structure decreases considerably its bandgap from about 7 to 3.6 eV [8]. However, the authors were uncapable of properly assigning the optical fingerprint of FL-phlogopite by standard Raman and FTIR spectroscopy experiments.

We present here SINS measurements on a FL-phlogopite flake to properly describe the phlogopite vibrational assignment at nanoscale. Fig. 6a shows the topography and IR broadband image of a selected staircase-like phlogopite flake exfoliated onto a 100 nm Au/Si. Fig. 6b compares the amplitude $S_2(\omega)$ and phase $\varphi_2(\omega)$ of the second harmonic point-spectra acquired at different phlogopite steps. The spectra revealed well-resolved IR absorption bands for FL-phlogopite located approximately at 472, 508, 682, 750, 809, 851, 960, 1001 and 1062 cm$^{-1}$. The assessment of these IR absorption bands observed through SINS technique can be done by comparing with previous micro-FTIR literature for bulk phlogopite [34,79]. The bands at 472 and 508 cm$^{-1}$ were assigned as Si-O vibrations; the band at 682 cm$^{-1}$ was assigned as Si-O-Mg vibration; the bands at 750, 809, and 851 cm$^{-1}$ were related to Al-O-Si stretching modes; while the bands at 960, 1001 and 1062 cm$^{-1}$ were assigned as Si-O-Si in plane stretching modes. We can conclude that phlogopite preserves its natural vibrational identity even when the material is thinned down to FLs.

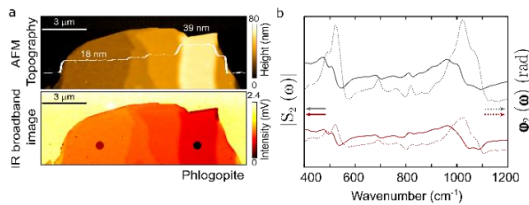

Fig. 6. Infrared Nanospectroscopy of phlogopite. (a) AFM topography and IR broadband image of staircase-like phlogopite flake exfoliated onto Au/Si substrate. (b) SINS point-spectra: amplitude $S_2(\omega)$ (continuous lines) and phase $\varphi_2(\omega)$ (dashed lines) of phlogopite flake with 18 nm (red) and 39 nm (black) of thickness acquired at the red and black circle positions in (a), respectively.

### C. Clinochlore - Chlorite

Clinochlore is a natural layered structured crystal with the chemical formula $Mg_5Al(AlSi_3)O_{10}(OH)_8$ that belongs to the chlorite group [1,2]. Besides being applied to diverse fields, such as prebiotic synthesis and polymerization of biomolecules[80], catalyst in the chemical recycling of waste plastics [81], decontamination of water resources [82], and hydration/weathering processes at nanoscale [83], it has recently been shown that clinochlore is an insulator (bandgap of ~3.6 eV) and can be mechanically exfoliated down to FL (Fig. 7a) [9]. Moreover, in this work de Oliveira and co-workers demonstrated that the spectral response and vibration signature should remain intact in exfoliated flakes. Fig. 7b brings a high-resolution AFM topography image (left panel), together with the broadband near-field image (right panel) acquired with sharp optical reflectivity contrast for different amounts of stacked layers (AFM height profile). Fig. 7c plots SINS point spectra of 11 and 20 nm clinochlore thick region (red and black dot position in Fig. 7b, respectively). The measured SINS spectra unveil a strong IR activity in the frequency range from 400-1150 cm$^{-1}$ that matches their micro-FTIR counterparts. Following the results obtained by micro-FTIR, the Si–O in-plane stretching vibration bands have been observed at 1023 and 987 cm$^{-1}$, while the Si–O in-plane bending vibration has been observed at 445 cm$^{-1}$. Al–O stretching is associated with the weak band at 811 cm$^{-1}$ and the two bands observed at 680 and 470 cm$^{-1}$ are assigned to the metal-OH bond libration and translation, respectively. In the 400–550 cm$^{-1}$ region, the IR spectra displayed three bands near 521, 492, and 420 cm$^{-1}$. These bands are assigned to (Fe,Mg)-O-Si or Al-O-Si bending [9].

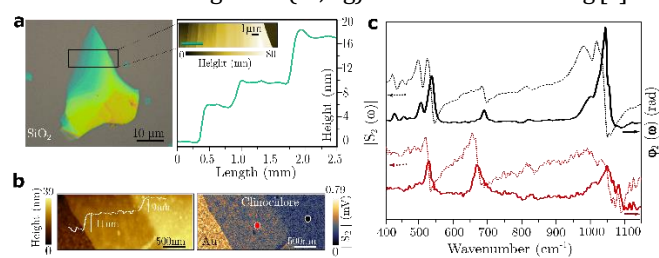

Fig. 7. Infrared Nanospectroscopy of Clinochlore. (a) Optical microscopy image of staircase-like clinochlore flake exfoliated onto $SiO_2$/Si substrate (left panel). AFM topography image (right panel) of the highlighted region as inserted in right and the corresponding profile along the green line in the AFM image. (b) AFM topography image of a FL-clinochlore flake onto a Au/Si substrate with respective AFM profile line (left panel) and broadband near field image of the same region revealing scattering intensity modulation according to the number of layers (right panel). (c) SINS point-spectra: amplitude $S_2(\omega)$ (dashed line) and phase $\varphi_2(\omega)$ (continuous line) of clinochlore flake with 11 nm (red) and 20 nm (black) of thickness acquired at the red and black dot position in (b), respectively. Adapted from Ref. [9].

### D. Talc – Clay

Talc, or also called soapstone, is an insulator (bandgap of ~5 eV) [3] and 2:1 layered magnesium hydrosilicate mineral with chemical formula $Mg_3Si_4O_{10}(OH)_2$. Talc is known as the softest mineral and shows perfect basal cleavage held together by van der Waals force [3,5,12,24,84]. This vdW crystal forms, by mechanical exfoliation, flat crystals with nanometer thicknesses, as observed with AFM topography (Fig. 8a). Exfoliated talc flakes have been used so far as substrates in ultrathin 2D-based devices [5,12,16], and in TMDs exciton physics experiments [16–18].

SINS measurements have been applied to ultrathin talc flakes to probe its vibrational fingerprints [24,48]. In Fig. 8a (bottom panel), SINS broadband image unveils sharp contrast among the talc layers and the $SiO_2$ substrate. Furthermore, Fig. 8b highlights the changes in the SINS spectra as a function of talc thickness. Fig. 8c shows IR resonances assigned to Si–O vibrations in the tetrahedral layer around $\nu_3$~990 cm$^{-1}$ and $\nu_2$~1010 cm$^{-1}$, as well as OH libration modes around $\nu_1$~670 cm$^{-1}$ and $\nu_{out}$ ~690 cm$^{-1}$. Considering the symmetry of those vibrations checked by first-principle calculations, $\nu_1$ and $\nu_3$ are in-plane modes, whilst $\nu_2$ and $\nu_{out}$ are out-of-plane vibrations. These results indicate that talc vibrational modes and their spectral response can be spectrally tuned upon controlling the number of layers and readily defining an IR fingerprint for the talc layers. Moreover, Ref. [24] has calculated the atomic displacement patterns of all IR active vibration modes of talc and correlate these motions to the intensity patterns in the SINS phase spectrum.

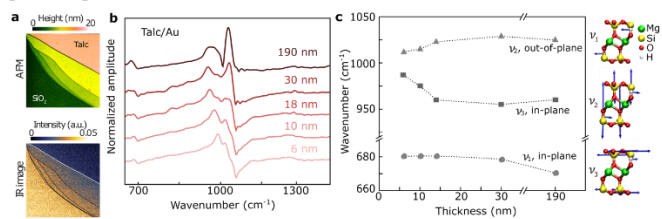

Fig. 8. Vibrational properties of Talc. (a) AFM topography and IR broadband nearfield image of a staircase-like talc on a $SiO_2$ substrate. (b) SINS point-spectra of talc nanocrystals with different layers on the Au/Si substrate. (c) Position of the maxima of the talc peaks, which are labeled according to the theoretical assignment (right panel), as a function of the talc thickness on the Au substrate. Si, Mg, O, and H atoms are represented by yellow, green, red, and white spheres, respectively. Figure adapted from Ref. [48]. Copyright 2018: The American Chemical Society.

### E. Antigorite – Serpentine

Natural obtained serpentines have theoretical formula $Mg_3Si_2O_5(OH)_4$, but can be arranged in three main structural forms of serpentine, classified based on their different T-$O_c$ stacking arrangements: lizardite (flat layers), chrysotile (curled layers) and antigorite (corrugated layers) [85]. In lizardite, the resulting structure is planar, owing to shifts of the $O_c$ and T cations away from their ideal positions and to the limited Al-for-Si substitution in the T sites. In antigorite, the misfit is compensated by a modulated wave-like structure, in which the sheet of $O_c$ is continuous and the tilted $SiO_4$ tetrahedra periodically switch their orientation, pointing alternatively in opposite directions. Chrysotile forms cylindrical or spiral tube structures, which are responsible for its properties as asbestos in insulating and fire-resistant materials [86]. Recent works have also shown by ab initio simulations that the serpentine class have a bandgap of about 4-4.2 eV, besides, obtain FL-antigorite flakes by mechanical exfoliation [11]. Fig, 9a then shows the AFM image of a representative staircase-like serpentine flake (in green-orange shades) deposited atop the $SiO_2$/Si substrate (in blue). The dashed black line in Fig. 9a indicates the position where a line profile of the serpentine flake was acquired and it resolves steps a ~1 nm-thick layer within the serpentine flake, suggesting that isolated serpentine monolayers could, eventually, be produced.

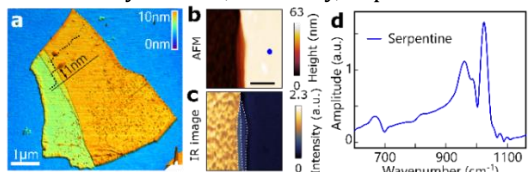

Fig. 9. Infrared Nanospectroscopy of Serpentine. (a) AFM topography image of an antigorite-serpentine flake (green-orange shades) atop the $SiO_2$/Si substrate (blue shades). (b) AFM topography and (c) Broadband IR image of serpentine multilayer in the same region, the scale bar represents 500 nm. (d) SINS point spectra of serpentine (blue curve) obtained at the location marked by a blue circle in (b). Figure adapted from Ref. [11].

Fig. 9b-d indicates that their structural signature remains intact, or, in other words, their serpentine nature is not modified during the exfoliation procedure. Fig. 9b-c presents the AFM topography (Fig. 9b) and broadband IR image (Fig. 9c) of a representative serpentine flake. The IR response map (Fig. 9c) is generated by the direct integration of the tip-scattered signal across the whole spectral sensitivity of the detector. Finally, Fig. 9d shows a standard SINS point spectrum obtained in the mid-IR range (600-1300 $cm^{-1}$) at the location marked in Fig. 9b. All measured spectra in this IR range show only Si-O vibrational modes. In this region, the authors described the strong vibrational responses at 960, 990 and 1025 $cm^{-1}$ to Si-O stretching modes, while the band at 666 $cm^{-1}$ is related to Si-O bending mode.

### F. Electrically tunable hybrid modes in graphene/talc heterostructures

As described in the previous sections, the phonon-polariton of talc can be easily probed by SINS. Moreover, graphene, another natural material, has shown very interesting optoelectronic properties, for instance, controlling light at the nanoscale [87]. These experiments were mainly carried out with s-SNOM and demonstrated the ability to confine and control light and manipulate it through regulation of graphene Fermi level [88–90]. So far, several works presented hybrid modes when graphene was deposited over different 2D substrates [91–93]. Thus recently, a natural atomically flat vdW heterostructure has been studied in a graphene/talc nanophotonic device and a hybrid plasmon phonon-polaritons were identified using SINS. In this work, Barcelos and colleagues demonstrated that the SINS spectra of talc vary as a function of its number of layers, but more important the hybrid plasmon phonon-polaritons in a graphene/talc heterostructure can be finely controlled by an external bias when both materials are assembled in a transistor-like architecture [48].

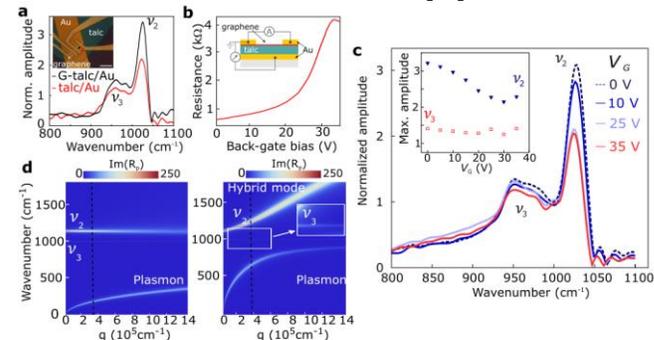

Fig. 10. Hybrid polaritons in Graphene/Talc heterostructure. (a) SINS point-spectra of the talc/Au (red curve) and G-talc/Au (black curve) were obtained at the locations marked by red and black dots in the optical image shown in the inset. (b) Resistance vs. back-gate voltage ($V_G$) for the G-talc/Au heterostructure. The inset illustrates the configuration for the G-talc/Au transistor measured. (c) SINS point-spectra for the G-talc/Au heterostructure for selected gate bias. In the inset are presented the maxima intensities of the SINS point-spectra for all gating measurements, demonstrating the gate tunability of the plasmon-phonon polaritons coupling between graphene and talc. (d) Hybridization of graphene plasmons and talc phonons: calculated dispersion relations for graphene-talc system for weakly (left panel, $n=1\times10^{10}\,cm^{-2}$) and moderate (right panel, $n=6.5\times10^{12}\,cm^{-2}$) doped graphene. The vertical dotted line indicates the momentum excited in by the SINS system due to the AFM tip radius. Figure adapted from Ref. [48]. Copyright 2018: The American Chemical Society.

Fig. 10a plots SINS point-spectra for regions on and off the graphene layer and one can note that the $\nu_2$=1025 $cm^{-1}$ out-of-plane mode is significantly increased for graphene/talc compared to pure talc. Whereas there is a slight modification for the $\nu_3$=960 $cm^{-1}$ mode. The transfer curve of the graphene/talc device (Fig. 10b) depicts the graphene charge neutrality at a back-gate voltage $V_G$~32 V, corroborating the spontaneous p-type doping of graphene at $V_G$=0 V [5]. By performing in situ SINS measurements as a function of $V_G$, a clear amplitude modulation of the $\upsilon_2=1025$ cm$^{-1}$ mode was observed (Fig. 10c). These findings indicated a clear coupling between graphene plasmons and the talc phonons leading to the formation of hybrid plasmon-phonon polaritons. This hybridization is further confirmed by $\omega$–$q$ dispersion calculation (Fig. 10d), where a blueshift of the hybrid mode in the $\omega$–$q$ dispersion of the graphene-talc system in the high p-doping condition (n≈6.5×10$^{12}$ cm$^{-2}$ corresponding to $V_G$=0 V) is shown in the right panel of Fig. 10d, in comparison with pure talc on the left panel. Thus, the authors demonstrated the potential use of natural crystals as well as their vdW heterostructures to be a key element in low-cost ultracompact devices for long-wavelength and polariton-based mid-infrared applications.

### G. Flexible and electrically tunable plasmons in graphene/mica heterostructures

Flexible plasmonic devices with electrical modulation are of great interest for diverse applications, such as flexible metamaterials and wearable sensors and photonic devices. However, the traditional flexible metal–polymer plasmonic structures suffer from a lack of electrical tunability. Then, Hu and colleagues experimentally demonstrated a flexible, electrically tunable, and strain-independent plasmons based on graphene–mica heterostructures [94]. They have shown a robust study for resonance frequency, quality factor, electrical tunability and lifetime of graphene plasmons for different bending radius and bending cycles. They demonstrated that the plasmons properties did not degrade at all even after 1000 bending cycles at a bending radius of 3mm. Moreover, the authors have used s-SNOM technique for assessing the graphene plasmon properties on SiO$_2$ and mica substrates.

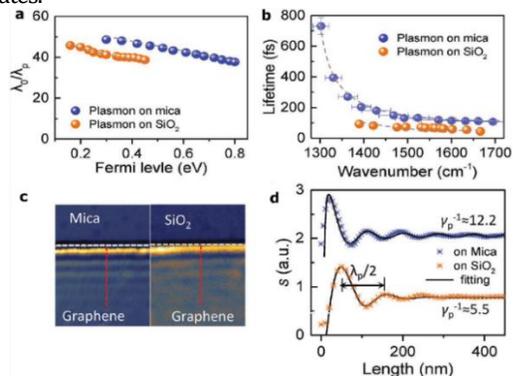

Fig. 11. Highly confined and long lifetime of flexible plasmon in a Graphene/Mica heterostructure. (a,b) Graphene plasmon confinement ($\lambda_0/\lambda_p$) and lifetime (t) as a function of Fermi level, respectively. (c) Near-field amplitude s($\omega$=895 cm$^{-1}$) images of graphene plasmon on mica and SiO$_2$ substrates (third-order demodulated harmonics of the near-field amplitude). (d) Line profiles of plasmon intensity across the graphene edge (white dash lines in (c)). Figure adapted from Ref. [94] under the terms of the CC BY 4.0 license.

Fig. 11a displays the field localization of plasmon in the flexible graphene/mica heterostructure device. Whereas Fig. 11b comparatively shows the plasmons lifetime of graphene over the mica and on SiO$_2$ substrates. The authors observed a strong IR-active phonon absorption from 950 to 1200 cm$^{-1}$ (1168 cm$^{-1}$) in the mica (SiO$_2$) substrate, which corroborated the strong plasmon–phonon coupling in the graphene/mica (graphene/SiO$_2$) devices. They then performed the real-space imaging of plasmon fields of the flexible graphene/mica devices. Fig. 11c shows representative near-field images on mica and SiO$_2$ substrates. The fringes parallel to the graphene edge are formed by the interference of tip-launched forward propagating plasmons and the partially reflected plasmon waves by the edge, and the oscillation period equals to $\lambda_p$/2. Fig. 11d plots line profiles of the near-field signals across the edge at the excitation wavelength of 895 cm$^{-1}$ ($\lambda_0$=11.12 µm). The authors have found a plasmon wavelength $\lambda_p$=196 nm (220 nm) for the graphene/mica heterostructure (graphene/SiO$_2$), corresponding to wavelength confinement of ~57 (~51), as well as an inverse damping ratio of 12.2 on graphene/mica heterostructures, which was approximately two times larger than that on SiO$_2$. Finally, the authors claim that their results provide the basis for the design of flexible active nanophotonic devices.

**Table 1** - Summary of the IR-active vibrational modes of phyllosilicates probed by nano-FTIR. The references for all mode assignments can be found in Section 4 for each phyllosilicate.

| Group | Specimen | Chemical formula | Frequency (cm$^{-1}$) | Assignment |
|---|---|---|---|---|
| Mica | Muscovite | KAl$_2$(Al,Si$_3$)O$_{10}$(OH)$_2$ | 1080 | Si-O stretching |
|  |  |  | 831 | Al-O stretching mixed |
|  |  |  | 920 | Al-OH/Al-O-Al stretching and Si-O-Si/Si-O-Al stretching |
|  | Phlogopite | KMg$_3$(AlSi$_3$)O$_{10}$(OH)$_2$ | 472 | ∥ Si-O stretching |
|  |  |  | 508 | ∥ Si-O stretching |
|  |  |  | 682 | Si-O-Mg stretching |
|  |  |  | 750 | ∥ Si/Al-O-Si stretching |
|  |  |  | 809 | ∥ Si/Al-O-Si stretching |
|  |  |  | 851 | ∥ Si/Al-O-Si stretching |
|  |  |  | 960 | ⊥ Si/Al-O stretching |
|  |  |  | 1001 | ∥ Si-O stretching |
|  |  |  | 1062 | ∥ Si-O stretching Si-O stretching |
| Chlorite | Clinochlore | Mg$_5$Al(AlSi$_3$)O$_{10}$(OH)$_8$ | 420 | (Fe, Mg)-O-Si or Al-O-Si bending |
|  |  |  | 445 | ∥ Si-O bending |
|  |  |  | 470 | metal-OH translation |
|  |  |  | 492 | (Fe, Mg)-O-Si or Al-O-Si bending |
|  |  |  | 521 | (Fe, Mg)-O-Si or Al-O-Si bending |
|  |  |  | 680 | metal-OH libration |
|  |  |  | 811 | Al-O stretching |
|  |  |  | 987 | ∥ Si-O stretching |
|  |  |  | 1023 | ∥ Si-O stretching |
| Clay | Talc | Mg$_3$Si$_4$O$_{10}$(OH)$_2$ | 670 | ∥ OH libration |
|  |  |  | 690 | OH libration |
|  |  |  | 990 | ∥ Si-O stretching |
|  |  |  | 1010 | ⊥ Si-O stretching |
| Serpentine | Antigorite | Mg$_3$Si$_2$O$_5$(OH)$_4$ | 666 | Si-O bending |
|  |  |  | 960 | Si-O stretching |
|  |  |  | 990 | Si-O stretching |
|  |  |  | 1025 | Si-O stretching |

## 5. SUMMARY AND OUTLOOK

This review paper brings an update on how SINS has approached the vibrational identity of naturally occurring vdW crystals, mainly belonging to the phyllosilicate class (Table 1). We have briefly described the structure of such 2D materials, the limitations of using Raman spectroscopy to assess their vibrational modes in the ultrathin limit, followed by the SINS experimental scheme and data collection. Due to its wide IR coverage and high sensitivity to Si-O vibrational activity, SINS is a decisive analytical tool for phyllosilicates, allowing monitoring their IR signatures from mid- to far-IR in a simultaneous experiment that connects morphology and nano-optics of these minerals in their atomic-thin form. The enhanced SNR of SINS enables the access to the chemistry of few layers of phyllosilicates in a non-destructive fashion, as it operates with an extreme low power IR beam. Our work then highlights the SINS contribution to the study of different phyllosilicates: micas, clays, chlorites, and serpentines, as well as the plasmonic tunability in graphene/talc and graphene/mica heterostructures.

**Acknowledgments.** The authors acknowledge financial support from the Brazilian Institute of Science and Technology (INCT) in Carbon Nanomaterials, and The National Council for Scientific and Technological Development (CNPq). The authors like to acknowledge the Brazilian Synchrotron Light Laboratory (LNLS), part of the Brazilian Centre for Research in Energy and Materials (CNPEM), a private non-profit organization under the supervision of the Brazilian Ministry for Science, Technology, and Innovations (MCTI). Also, the Advanced Light Source (ALS) for the facilities in experiments involving synchrotron radiation and its associated installations, besides Hans A. Bechtel and Stephanie G. Corder for the experimental assistance. The authors thank the Brazilian Nanotechnology National Laboratory (LNNano – Proposals No. 20230143, 20221047, 20221050, and 20221696), the Microscopic Samples Laboratory (LAM – Proposals No. 20221265 and 20221266), and the IMUBIA beamline (Proposals No. 20220423 and 20220584) at CNPEM for sample fabrication and characterization. R.O. thanks the Coordination for the Improvement of Higher Education Personnel (CAPES) and research supporting foundation of Minas Gerais state (FAPEMIG). I.D.B., A.R.C. and R.O.F. acknowledge the support from CNPq through the Research Grants 311327/2020-6, 309920/2021-3, and 309946/2021-2. I.D.B acknowledges the prize L'ORÉAL-UNESCO-ABC for Women in Science Prize - Brazil/2021. R.O.F acknowledges the FAPESP financial support (Young Investigator Grant 2019/14017-9).

**Disclosures**. The authors declare no conflicts of interest.
**Data Availability.** Data underlying the results presented in this paper are not publicly available at this time but may be obtained from the authors upon reasonable request.